# Are Chlorine Isotopologues of Polychlorinated Organic Compounds Exactly Binomially Distributed? A Theoretical Study and Implications to Experiments

**Caiming Tang**[1,2,*]

[1] *State Key Laboratory of Organic Geochemistry, Guangzhou Institute of Geochemistry, Chinese Academy of Sciences, Guangzhou 510640, China*

[2] *University of Chinese Academy of Sciences, Beijing 100049, China*

*Corresponding Author.

Tel: +86-020-85291489; Fax: +86-020-85290009; E-mail: CaimingTang@gig.ac.cn. (C. Tang).

Preprint version submitted to ArXiv.org.

# ABSTRACT

Chlorine isotopologues of polychlorinated organic compounds are usually recognized to follow binomial distribution. Is this recognition exactly true? This study presents a solid theoretical derivation to prove whether the isotopologue distributions of organochlorines are binomial or not, and investigates the implications of the distributions to relevant experimental studies. The fundamental principles causing different chlorine kinetic isotope effects (KIEs) were discussed. During synthetic reactions, the C-Cl bonds in possession of stronger strengths are more readily to be formed with heavy isotopes. The chlorine KIEs are higher during the breakages of the stronger C-Cl bonds than during the breaking of the weaker ones. For a synthetic organochlorine, if the rate-limiting step during a chlorination reaction is the formation of C-Cl bonds of the organochlorine, then chlorine KIE takes effect in the reaction, resulting in always higher chlorine isotope ratio on the reaction position than that of the initial chlorine atoms from the chlorine source. Chlorine KIEs vary during different chlorination reactions, leading to inconsistent chlorine isotope ratios on different reaction positions of asymmetric reaction intermediates and final products. If the rate-determining step is the generation of reactive chlorine atoms and the initial isotope ratio of the chlorine source is constant, then the chlorine isotope ratios on the reaction positions of all intermediates and final products are equivalent, demonstrating that the isotopologues follow binomial distribution. After physical changes and dechlorination reactions, organochlorines in the environment are unlikely binomially distributed. Experimental data show that the detected isotopologues of all the investigated organochlorines on electron ionization mass spectrometry disobeyed binomial distribution. The inconsistent isotopologue distributions may trigger deviations in quantification and compound-specific chlorine isotope analysis (CSIA-Cl) of organochlorines. Using more than one highest-abundance isotopologue for quantification and application of complete-isotopologue scheme of isotope ratio calculation for CSIA-Cl are proposed for achieving high-quality data. Gas chromatography-double focus magnetic-sector high resolution mass spectrometry may be a promising tool for CSIA-Cl using the complete-isotopologue scheme, due to its excellent performances in sensitivity and selectivity. This study may bring people a new perspective about the chlorine isotopologue distributions of organochlorines, and the proposed solutions may help to obtain better experimental results for quantitative analysis and CSIA-Cl of organochlorines.

**Keywords:**

# INTRODUCTION

Chlorinated organic compounds, which can be produced in both anthropic and natural activities, are ubiquitously present in the world. Anthropogenic organochlorines are numerous, such as chloroethylenes, polyvinyl chloride, chlorinated hydrocarbons, Cl-containing pharmaceuticals and etc. In addition, naturally occurring organochlorines are a variety of compounds such as chloramphenicol, aureomycin, vancomycin and etc. So far, around 3800 halogen-containing natural products have been found, most of which are chlorinated/brominated and few are fluorinated/iodinated (Gribble, 2003). Chlorinated organic compounds have significantly changed the world along with human being live, and have been affecting many aspects of human activities.

Organochlorines are a very important category of environmental organic pollutants, such as chlorofluorocarbons, chloroethylenes, chlorinated hydrocarbons and the chlorinated persistent organic pollutants (POPs) as listed in the Stockholm Convention on POPs including dichlorodiphenyltrichloroethane (DDT), dioxins, hexachlorobenzene, polychlorinated biphenyls and so on (United Nations Environment Programme, 2015). These pollutants have caused serious environmental pollution and posed disastrous risks to human heathy and ecosystem globally (Carson, 2002; Longnecker et al., 1997).

Chlorine has two natural stable isotopes ($^{35}Cl$ and $^{37}Cl$) with a globally average ratio ($^{35}Cl/^{37}Cl$) of 0.7578:0.2422 (Hoefs, 2015), thus organochlorines have specific isotopologue distributions (Pena-Abaurrea et al., 2014). Usually, the isotopologue distributions of organochlorines are roughly considered as binomial (Anderegg, 1981; Sakaguchi-Söder et al., 2007). However, do they really exactly comply with binomial distribution? So far, no available report has confirmed this issue. In many research fields, quantification of organochlorines is of important significance. Analysts always tacitly approve that the isotopologues of organochlorines are binomial distributed and the distribution shapes are negligibly different, and accordingly generally apply only one isotopologue (usually the theoretically most abundant one) to the quantification when using mass spectrometry-related techniques. No study has questioned the validity of this quantification manner. No available data have shown that isotopologue distributions can trigger analytical errors and to what extent the errors are negligible. In environmental study areas, compound-specific chlorine isotope analysis (CSIA-Cl) has been becoming increasingly important during the last two decades (Schmidt et al., 2012). However, the currently available relevant studies focused on the overall chlorine isotope ratios only (Cincinelli et al., 2012), rather than the detailed information about the distributions of chlorine isotopologues of organochlorines. The validity of this approach to the source apportionment of organochlorines is somewhat ambiguous, due to possible misjudged results. For example, the samples with an equal overall chlorine isotope ratio may not stem from the same source, because their isotopologue distributions may be different.

Therefore, in the present study, we provides solid theoretical derivations for the causes of different kinetic isotope effects (KIEs), the chlorine isotopologue distributions of organochlorines after synthetic reactions, physical changes and dechlorination reactions, and how the isotopologue distributions affect the quantification and CSIA-Cl of organochlorines. Based on the experiment data, we evaluated the quantification errors caused by the different isotopologue distributions taking a trichlorobiphenyl (PCB-18) for an example. The changes of isotopologue distributions of organochlorines after dechlorination reactions and physical fractionation were investigated on the basis of the obtained experimental data of perchloroethylene (PCE), trichloroethylene (TCE) and PCB-18. Some suggestions for conducting CSIA-Cl studies were proposed, including the selections of external isotopic standards, mass spectrometry types and isotope ratio calculation schemes. The theoretical inferences and experimental proofs may bring a new perspective to people in terms of the distributions of chlorine isotopologues of organochlorines, and the solutions proposed in this study may be conducive to the quantitative analysis and CSIA-Cl of organochlorines for achieving high-quality data.

# PART 1: FUNDAMENTALS OF ISOTOPE EFFECTS

**Isotope Effects in Chemical Changes**

Two atoms and the linking chemical bond can be regarded as a simple harmonic vibration system, and the vibrational frequency $v$ is:

$$v = \frac{1}{2\pi}\sqrt{\frac{k}{\mu}} \quad (1.1)$$

where k is the force constant indicating the bond strength (like the spring's stiffness) and μ is the reduced mass of the two atoms and given by (Chacko et al., 2001):

$$\mu = \frac{m_1 m_2}{m_1 + m_2} \quad (1.2)$$

The vibrational energies of this vibration system are quantized and can be expressed as:

$$E = (n + 1/2)hv \quad (1.3)$$

where n denotes energy level, and h is the Planck constant. At 0 K, the molecule is in ground state (n=0). In this state, the zero point energy (ZPE) is not zero and given by:

$$ZPE = 1/2\,hv = \frac{h}{4\pi}\sqrt{\frac{k}{\mu}} \quad (1.4)$$

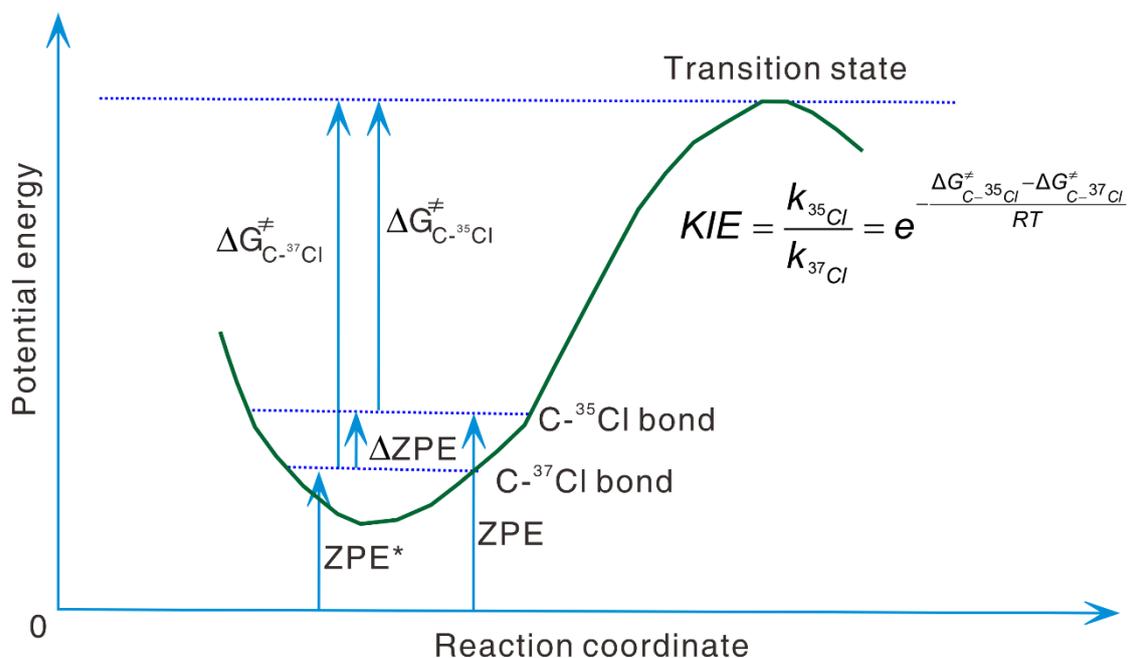

**Figure 1**. Energy level diagram of reactions involving chlorine isotopes. The zero point energy of a molecule formed by the heavy isotope (ZPE*) is lower than that of the molecule formed with the light one. The extent of ΔZPE (ZPE-ZPE*) for a compound play a major role on the kinetic isotope effect (KIE). The chemical bonds containing heavy isotopes are faster to be formed but slower to be cleaved than those containing light isotopes. $\Delta G^{\neq}$ represents the critical energies; R is the universal gas constant (0.00199 kcal/mol·K); T is temperature (K).

The bonds comprised of different isotopes have different vibrational frequencies, accordingly leading to different ZPEs.

Based on eqs 1.1 and 1.2, the vibrational frequencies are related to the corresponding reduced masses (μ and μ*) and can be expressed as:

$$\frac{v^*}{v} = \sqrt{\frac{\mu}{\mu^*}} \qquad (1.5)$$

where asterisk-marked parameters are related to the system involving heavy isotopes. From eq 1.5, it can be seen that the bonds involving heavy isotopes have lower vibrational energies than that with light isotopes, and the ZPEs of the heavy-isotope bonds are thus lower than those of the light-isotope bonds. The lower ZPEs suggest that the heavy-isotope bonds are more stable and can be formed with priority. Thus, the reactions involving heavy isotopes are faster than those involving light isotopes. In low temperature systems, the only difference between the heavy-isotope and light-isotope bonds are the difference between the ZPEs (ΔZPE), which can be expressed as:

$$\Delta ZPE = ZPE - ZPE^* = 1/2\, h(v - v^*) = 1/2\, h\Delta v \qquad (1.6)$$

In synthetic reactions, when relatively high ΔZPEs present, the heavy-isotope bonds are more readily formed. However, as temperature increase, the bond energy levels arise, the effects of ΔZPE are reduced, thus declining the differences of formation rates between the heavy-isotope and light-isotope bonds and reducing isotope effects.

Transforming eq 1.5, gives rise to:

$$\Delta v = v - v^* = v\left(1 - \sqrt{\frac{\mu}{\mu^*}}\right) \qquad (1.7)$$

where $\Delta v$ is the difference of the vibrational frequencies between the heavy-isotope and light-isotope bonds. It can be seen from this equation that the bonds with higher vibrational frequencies exhibit larger vibrational frequency differences ($\Delta v$) between the heavy-isotope and light-isotope bonds than those with lower vibrational frequencies. In other words, according to eqs 1.4 and 1.7, the stronger bonds have larger vibrational frequency differences between the heavy-isotope and light-isotope bonds than the weaker bonds.

According to eqs 1.6 and 1.7, the stronger bonds are more likely formed with heavy isotopes, resulting in more significant isotope effects in synthetic reactions in comparison with the weaker bonds (Figure 1). As a result, the bonds with different strengths may lead to different isotope effects during synthetic reactions.

To the contrary, during bond breakages, the chemical bonds with light isotopes are easier to break than those with heavy isotopes. As bond strengths increase, the differences of bond breaking rates between the heavy-isotope and light-isotope bonds become larger. In other words, bond breaking KIEs are in positive correlation with bond strengths (Figure 1).

**Isotope Effects in Physical Changes**

Isotope effects can take place during physical changes such as dissolution, evaporation, diffusion, adsorption, desorption and etc. Heavy-isotope bonds are shorter than light-isotope bonds, thus molecules containing heavy isotopes are more compact than those containing light isotopes. Smaller molecular volume can result in smaller polarizability, which leads to weaker intermolecular interactions. Thus, the heavier isotopologues have lower polarities than the lighter ones. Polarity is a key factor affecting behaviors of compounds during the processes of dissolution, evaporation, adsorption and desorption. Different polarities may lead to different changing rates and extents of isotopologues during physical changes. Analogously, due to the differences in polarity, the heavier and lighter isotopologues can show different behaviors during physical changes, leading to occurrences of isotope effects.

Diffusive isotope effects ($\varepsilon_{diff-air}$) take place during diffusion processes. Diffusion effect is mass-dependent and depends of molecular-weight differences among different isotopologues. The fractionation factor ($\alpha_{diff-He}$) stemming from the diffusion effect is calculated as (Craig, 1953):

$$\alpha_{diff-air} = \sqrt{\frac{M_l(M_h + M_{air})}{M_h(M_l + M_{air})}} \qquad (1.8)$$

where $M_l$ and $M_h$ denote to molecular weights of the lighter and heavier isotopologues, respectively and $M_{air}$ is the average molecular weight of air. Due to $M_h > M_l$, $\alpha_{diff-air}$ is always less than 1.

The diffusive isotope effects can then be given as (Mook et al., 2000):

$$\varepsilon_{diff-air} = f(1-s)(\alpha_{diff-air} - 1) \qquad (1.9)$$

where f is a correction factor for air flow ranging from 1 to 0.5, and s is related to the relative vapor saturation of target compounds (Mook et al., 2000). As can be seen from eq 1.9, diffusive isotope effects are always negative, indicating that the isotope effects during diffusion processes are always normal, namely, the lighter isotopologues diffuse faster than the heavier ones.

# PART 2: ISOTOPOLOGUE DISTRIBUTIONS OF SYNTHETIC ORGANOCHLORINES

**Symmetric Organochlorines**

Due to that reactions involving the formation or breaking of multiple bonds cannot normally be synchronous or concerted (Dewar, 1984), the multiple chlorination reactions of polychlorinated organic compounds are thus anticipated to be stepwise. In synthetic reactions of organochlorines, even though reactants and final products are symmetric compounds, the bond energies of the stepwise formed C-Cl bonds are different, because the C-Cl bond energies of the reaction intermediates with different numbers of Cl atoms are inconsistent. For instance, the bond energies of the C-Cl bonds of monochloromethane, dichloromethane, chloroform and tetrachloromethane are 350.2, 338.0, 311.1 and 296.6 kJ/mol at 298 K, respectively (Luo et al., 2012). As indicated in this case and the rest relevant cases in the literature, the C-halogen bond energies gradually decline along with the increase of the halogenation extents of halogenated congeners. Consequently, the KIEs during the stepwise chlorination reactions are likely different. If the rate-determining step of a chlorination reaction is the formation of C-Cl bonds, and the chlorine atoms (or other reactive forms) is constant or excessive, then chlorine KIE works during this reaction, leading to always higher chlorine isotope ratio on the reaction position than that of the initial chlorine atoms from the chlorine source. In this scenario, the KIEs during different chlorination reactions are different, leading to different chlorine isotope ratios on the different reaction positions of asymmetric reaction intermediates.

We hypothesize that the initial isotope ratio from the chlorine source ($IR_0$) is constant and expressed as:

$$IR_0 = \frac{^{37}Cl_0}{^{35}Cl_0} = \frac{b_0}{a_0} \qquad (2.1)$$

where a and b represent the relative abundances of $^{35}Cl$ and $^{37}Cl$, respectively.

Because the reaction rates of chlorination reactions involving $^{37}Cl$ are higher than those involving $^{35}Cl$, in other words,

$$KIE = \frac{k(^{35}Cl)}{k(^{37}Cl)} < 1 \qquad (2.2)$$

therefore, in the formed C-Cl bond i, we have

$$IR_i = \frac{^{37}Cl_i}{^{35}Cl_i} = \frac{b_i}{a_i} > \frac{b_0}{a_0} \qquad (2.3)$$

Due to the different KIEs during the stepwise chlorination reactions, the isotope ratios of different C-Cl bonds vary. Then, we have

$$\frac{b_i}{a_i} \neq \frac{b_j}{a_j} \quad (i \neq j) \qquad (2.4)$$

where i and j correspond to the C-Cl bonds i and j, respectively.

For a polychlorinated organic compound with different KIEs during stepwise chlorination reactions, the distribution function of the isotopologues can be given by:

$$g(n) = \prod_{i=1}^{n}(a_i + b_i) \qquad (2.5)$$

where n is the number of Cl atoms of the compound.

For the isotopologue containing $^{35}Cl$ only, the relative abundance is:

$$A_{^{35}Cl_n} = \prod_{i=1}^{n} a_i \qquad (2.6)$$

While for the isotopologue containing $^{37}Cl$ only, its relative abundance is:

$$A_{^{37}Cl_n} = \prod_{i=1}^{n} b_i \qquad (2.7)$$

And the overall isotope ratio can be calculated as:

$$IR = \frac{\sum_{i=1}^{n} b_i}{\sum_{i=1}^{n} a_i} \quad (2.8)$$

If the isotopologues comply with binomial distribution, then the following three equations can be received:

$$h(n) = \left(\frac{1}{n}\sum_{i=1}^{n} a_i + \frac{1}{n}\sum_{i=1}^{n} b_i\right)^n \quad (2.9)$$

$$\sqrt[n]{\prod_{i=1}^{n} a_i} = \frac{1}{n}\sum_{i=1}^{n} a_i \quad (2.10)$$

$$\sqrt[n]{\prod_{i=1}^{n} b_i} = \frac{1}{n}\sum_{i=1}^{n} b_i \quad (2.11)$$

We define a function k(x) as:

$$k(x) = \ln x \quad (2.12)$$

and then we have

$$\frac{d^2 k(x)}{dx^2} = -\frac{1}{x^2} < 0 \quad (2.13)$$

Therefore, k(x) is a concave function. According to the Jensen's inequality, we receive

$$\frac{1}{n}\sum_{i=1}^{n} \ln a_i \leq \ln\left(\frac{1}{n}\sum_{i=1}^{n} a_i\right) \quad (2.14)$$

which is equivalent to

$$\ln\left(\prod_{i=1}^{n} a_i\right)^{\frac{1}{n}} \leq \ln\left(\frac{1}{n}\sum_{i=1}^{n} a_i\right) \quad (2.15)$$

Thus, we have

$$\sqrt[n]{\prod_{i=1}^{n} a_i} \leq \frac{1}{n} \sum_{i=1}^{n} a_i \qquad (2.16)$$

And only when $a_1 = a_2 = ... = a_n$, the following equation can be obtained:

$$\sqrt[n]{\prod_{i=1}^{n} a_i} = \frac{1}{n} \sum_{i=1}^{n} a_i \qquad (2.17)$$

Thus, when $a_i \neq a_j \ (i \neq j)$, we have

$$\sqrt[n]{\prod_{i=1}^{n} a_i} < \frac{1}{n} \sum_{i=1}^{n} a_i \qquad (2.18)$$

Then the eqs 2.10 and 2.11 are disconfirmed, indicating that the isotopologues disobey binomial distribution.

If the rate-determining step is the production process of the reactive chlorine atoms, then the reactive chlorine atoms are almost completely consumed to form C-Cl bonds just after their generation, leading to the chlorine isotope ratio on the reaction position equivalent to that of the chlorine source, provided the initial isotope ratio of the chlorine source is constant. In this case, the chlorine isotope ratios on all the reaction positions of the intermediates and final product are identical, suggesting that the isotopologues follow binomial distribution.

**Asymmetric Organochlorines**

For asymmetric polychlorinated organic compounds, the KIEs during different chlorination reactions are more likely different in comparison with symmetric organochlorines. However, whether the isotopologues follow binomial distribution or not is determined by the rate-limiting step of the chlorination reactions. The scenarios for asymmetric organochlorines are analogous to the cases for symmetric organochlorines with regard to the distributions of isotopologues.

# PART 3: ISOTOPOLOGUE DISTRIBUTIONS OF ENVIRONMENTAL ORGANOCHLORINES

**After Physical Changes**

Physical changes such as dissolution, evaporation, diffusion, adsorption, desorption and etc. may lead to isotope effects. During these processes, organochlorine molecules are separated into at least two parts or phases, and the isotopologue compositions become different in different parts due to isotope effects. As a result, the chlorine isotope ratios in different parts are different, and certainly inconsistent with the initial isotope ratio before separation.

We hypothesize an organochlorine whose isotopologues comply with binomial distribution, and its chlorine isotope ratio (IR) is

$$IR = \frac{^{37}Cl}{^{35}Cl} = \frac{b}{a} \quad (a+b=1) \quad (3.1)$$

Then the distribution function of isotopologues can be expressed as:

$$f(n) = (a+b)^n \quad (3.2)$$

where n is the number of Cl atoms of the organochlorine. Then the relative abundance of the isotopologue containing i $^{37}Cl$ is

$$A_i = C_n^i a^{n-i} b^i \quad (3.2)$$

Thus the isotope ratio can be calculated with the relative abundances of any pair of adjacent isotopologues as follows (Elsner, 2008; Jin 2011):

$$IR = \frac{i}{n-i+1} \cdot \frac{A_i}{A_{i-1}} = \frac{b}{a} \quad (3.4)$$

And the overall isotope ratio calculated with the complete isotopologues ($IR_{comp\text{-}iso}$) of the organochlorine is:

$$IR_{comp\_iso} = \frac{\sum_{i=0}^{n} iA_i}{\sum_{i=0}^{n}(n-i)A_i} = \frac{b}{a} \quad (3.5)$$

Therefore, we have

$$\frac{i}{n-i+1} \cdot \frac{A_i}{A_{i-1}} = \frac{\sum_{i=0}^{n} iA_i}{\sum_{i=0}^{n}(n-i)A_i} \quad (3.6)$$

We hypothesize a certain molar amount of an organochlorine of which the isotopologues follow binomial distribution and the initial chlorine isotope ratio before separating into two groups is

$$IR_0 = \frac{b_0}{a_0} \quad (3.7)$$

And the chlorine isotope ratios of the two groups are

$$IR_I = \frac{b_I}{a_I} \quad (3.8)$$

and

$$IR_{II} = \frac{b_{II}}{a_{II}} \quad (3.9)$$

respectively.

We hypothesize that the abundances of the initial isotopologues were $A_{00}, A_{01}...A_{0i}...A_{0n}$, and those of the isotopologues of the separated two groups were $A_{I0}, A_{I1}...A_{Ii}...A_{In}$, and $A_{II0}, A_{II1}...A_{IIi}...A_{IIn}$, respectively.

Then we have

$$IR_0 = \frac{i}{n-i+1} \cdot \frac{A_{0i}}{A_{0(i-1)}} = \frac{b_0}{a_0} \quad (3.10)$$

We hypothesize that the isotopologues of a group (I) still follow binomial distribution, then receive

$$IR_I = \frac{i}{n-i+1} \cdot \frac{A_{Ii}}{A_{I(i-1)}} = \frac{b_I}{a_I} \quad (3.11)$$

Then the ratios of the isotopologue abundances of group I relative to the initial abundances can be given by:

$$a_1 = \frac{A_{I(1-1)}}{A_{0(1-1)}} \ldots a_i = \frac{A_{I(i-1)}}{A_{0(i-1)}} \ldots a_{n+1} = \frac{A_{In}}{A_{0n}} \quad (3.12)$$

Then for a random pair of neighboring isotopologues with i-1 and i $^{37}$Cl atoms, respectively, according to eqs 3.10 and 3.11, we have

$$IR_I = \frac{a_{i+1} A_{0i}}{a_i A_{0(i-1)}} \cdot \frac{i}{n-i+1} \quad (3.13)$$

which transforms to

$$\frac{A_{0i}}{A_{0(i-1)}} \cdot \frac{i}{n-i+1} = \frac{a_i}{a_{i+1}} IR_I = IR_0 \quad (3.14)$$

Since the isotopologues comply with binomial distribution, then the isotope ratios calculated with random different pairs of adjacent isotopologues are equivalent. Thus, for the isotope ratios calculated with two random adjacent pairs of neighboring isotopologues (three random neighboring isotopologues), we have

$$\frac{a_i}{a_{i+1}} IR_I = \frac{a_{i+1}}{a_{i+1+1}} IR_I \quad (3.15)$$

which simplifies to

$$\frac{a_i}{a_{i+1}} = \frac{a_{i+1}}{a_{i+1+1}} \quad (3.16)$$

Therefore, the progression ($a_1$, $a_2$ … $a_i$, $a_{i+1}$ … $a_n$, $a_{n+1}$) is geometric:

$$a_i = a_1 q^{i-1} \qquad (3.17)$$

and the common ratio (q) is:

$$q = \frac{a_2}{a_1} = \frac{a_{i+1}}{a_i} \qquad (3.18)$$

Thus, if the isotopologues follow binomial distribution, then the progression ($a_1$, $a_2$ … $a_i$, $a_{i+1}$ … $a_n$, $a_{n+1}$) must be geometric.

Due to the isotope effects in the separation, the chlorine isotope ratio of the isotopologues in group I become different from the initial isotope ratio. We hypothesize

$$IR_I = \frac{b_I}{a_I} > IR_0 = \frac{b_0}{a_0} \qquad (3.19)$$

which means the heavier isotopologues are enriched in group I and reduced in group II. Thus, the ratio ($a_i$) for a lighter isotopologue is lower than that for a heavier one ($a_{i+1}$). Thus, q is higher than 1 ($q > 1$).

If $n \to \infty$, the limit of $a_i$ is

$$\lim_{i \to n} a_i = \lim_{i \to \infty} a_1 q^{i-1} = \infty > 1 \qquad (3.20)$$

which is obviously impossible. Therefore, in fact, the progression ($a_1$, $a_2$ … $a_i$, $a_{i+1}$ … $a_n$, $a_{n+1}$) should not be geometric, and the isotopologues of group I should not comply with binomial distribution.

Actually, the maximum of $a_{n+1}$ is can only approach to 1, or in other words:

$$\lim_{n \to \infty} a_{n+1} = \lim_{n \to \infty} \frac{A_{In}}{A_{0n}} = 1 \qquad (3.21)$$

which is also unlikely in reality, due to that the heavier isotopologues cannot completely stay in only one group.

If the heavier isotopologues are reduced in group I and enriched in group II. Thus, the ratio ($a_i$) for a lighter isotopologue is higher than that for a heavier one ($a_{i+1}$). Thus, q is less than 1 ($q<1$).

If $n \to \infty$, the limit of $a_i$ is

$$\lim_{i \to n} a_i = \lim_{i \to \infty} a_1 q^{i-1} = 0 \quad (3.20)$$

which is also impossible. Therefore, in fact, the progression ($a_1$, $a_2$ … $a_i$, $a_{i+1}$ … $a_n$, $a_{n+1}$) should never be geometric. We thus deduce that the isotopologues of group I do not comply with binomial distribution.

Only in the case of $q=1$, namely, no isotope effect presents in the separation process, then the isotopologues of the two separated groups of the organochlorine can follow binomial distribution. For example, direct separation of one bottle of the organochlorine into two bottles may be in this scenario.

The above deduction is based on the prerequisite that the chlorine isotopologues of the organochlorine follow binomial distribution. If the initial isotopologues are not binomially distributed, how does the initial isotopologue distribution affect those in the two separated groups? Obviously, if the initial isotopologues disobey binomial distribution, the isotopologues in the separated groups are extremely impossible to comply with binomial distribution, and the initial isotopologues to some extent can impact the separated isotopologues in terms of distributions.

If a group of an organochlorine is separated into more than two groups and chlorine isotope effects take effect in the separation processes, it can be deduced that the separated isotopologues are not binomially distributed.

Besides separation, mixing process of different groups of an organochlorine should also be taken into account when investigating the distributions of isotopologues. We hypothesize two groups of an organochlorine of which the isotopologues comply with binomial distribution. If the two groups of the organochlorine are mixed, do the mixed isotopologues still follow binomial distribution?

This proposition can be proved with a proof by contradiction referencing to the case of separation as provided above. If the combined isotopologues still comply with binomial distribution, then the combined group of the organochlorine definitely can be separated back to the two original groups before combining. As proved above, only when $q=1$, can one group of an organochlorine of which the isotopologues are binomial distributed be separated into two groups in both of which the isotopologues follow binomial distribution. Therefore, only when the isotope ratios of the two initial groups of the organochlorine are equal, can the combined group have the isotopologues following binomial distribution.

If the isotopologues in the two groups are not binomially distributed, then the combined isotopologues cannot comply with binominal distribution statistically.

If more than two groups of an organochlorine are combined and the initial chlorine isotope ratios are different, it can be deduced that the combined isotopologues are not binomially distributed.

**After Dechlorination Reactions**

The dechlorination reactions of polychlorinated organic compounds in the environment may be analogous to those taking place in electron ionization mass spectrometry (EI-MS). Therefore, after environmental dechlorination reactions, the remaining parent molecules of an organochlorine may be similar to the detected molecular ions of the organochlorine undergoing dechlorination fragmentations in EI-MS. As proved in a previous study, the isotopologues of the detected molecular ions cannot obey binomial distribution, no matter whether the initial isotopologues of the compound follow binomial distribution or not (Tang et al., 2017a). Therefore, the isotopologues of the remaining fraction of an organochlorine after environmental dechlorination reactions can be inferred to disobey binomial distribution, regardless of that the initial isotopologues obey or disobey binomial distribution. For dechlorination products, due to further dechlorination reactions, their isotopologues cannot follow binomial distribution neither.

## PART 4: IMPLICATIONS TO ANALYTICAL STUDIES

**Quantification of Organochlorines**

As deduced above, the distributions of isotopologues of an organochlorine from different sources are always inconsistent. In present, almost all the quantitative analysis methods for organochlorines using mass spectrometric detection use only one isotopologue ion (always the highest-abundance ion) of individual compound. Are these analytical methods exactly rational? To what extent do these methods reveal the actual concentrations of the analyzed organochlorines?

We hypothesize that an organochlorine in a real sample has different isotopologue distribution from that of this compound in a calibration standard, and the isotopologue ion selected for quantification has the mass spectrometric signal intensities $I_1$ and $I_2$ in the real sample and in the calibration standard, respectively. If the concentration of the compound in the standard is $C_1$, then the concentration in the real sample ($C_2$) calculated with the conventional methods is

$$C_2 = C_1 \frac{I_2}{I_2} \qquad (4.1)$$

If the relative abundances of the quantification isotopologue in the calibration standard and in the real sample are $A_1$ and $A_2$, respectively, then we have:

$$C_2^* A_2 = C_1 A_1 \frac{I_2}{I_2} \qquad (4.2)$$

where $C_2^*$ is the actual concentration of the organochlorine in the real sample. Therefore, the quantification bias ($\Delta C$) of the measured concentration to the actual concentration can be given by:

$$\Delta C = \frac{C_2 - C_2^*}{C_2^*} = \frac{A_2}{A_1} - 1 = \Delta A \qquad (4.3)$$

Where $\Delta A$ is the relative deference between $A_1$ and $A_2$.

In reality, the absolute value of $\Delta A$ is anticipated to be very small and positively correlated to the absolute value of the relative deference ($\Lambda^{37}Cl$) between the chlorine isotope ratios of the organochlorine in the calibration standard and in the real sample. For instance, we found a very large chlorine isotope effect of 2,2'5-trichloro-1,1'biphenyl (PCB-18) on a gas chromatographic column with the $\Lambda^{37}Cl$ of 73.1‰ between the first and the last retention time segments (Tang et al., 2017b). We regard the PCB-18 detected in the first retention-time segment as the calibration standard and that detected in the last one as the analyte in the real sample. Then the quantification bias calculated by the theoretically highest-abundance isotopologue ($C_{12}H_7{}^{35}Cl_3$) is 5.2%, which may not be a negligible deviation in some strict quantification studies. While using the theoretically second highest-abundance isotopologue ($C_{12}H_7{}^{35}Cl_2{}^{37}Cl$) or the combination of the first two highest-abundance isotopologues, the quantification biases are reduced to -1.6% and 1.6%, respectively. We thus propose using two or more highest-abundance isotopologues to quantify an organochlorine for reducing quantification bias, although it is anticipated to be always very small.

**Table 1**. Quantification biases calculated with the measured relative abundances of different isotopologues of PCB-18 in the first and the last retention-time segments.

| Isotopologue formula | Theoretical relative abundance | First retention segment | | Last retention segment | | Quantification bias | Quantification bias* |
|---|---|---|---|---|---|---|---|
| | | Relative abundance | SD (1σ, n=5) | Relative abundance | SD (1σ, n=5) | | |
| $C_{12}H_7{}^{35}Cl_3$ | 0.43499 | 0.39330 | 0.00169 | 0.41393 | 0.00528 | 5.2% | 1.6% |
| $C_{12}H_7{}^{35}Cl_2{}^{37}Cl$ | 0.41738 | 0.44504 | 0.00288 | 0.43789 | 0.00233 | -1.6% | |
| $C_{12}H_7{}^{35}Cl{}^{37}Cl_2$ | 0.13341 | 0.14605 | 0.00209 | 0.13444 | 0.00323 | -7.9% | |
| $C_{12}H_7{}^{37}Cl_3$ | 0.01422 | 0.01562 | 0.00029 | 0.01373 | 0.00094 | -12.1% | |

Note, *: the bias calculated with the combination of the first two highest-abundance isotopologues.

**CSIA-Cl of Organochlorines**

Due to various physical and chemical changes of organochlorines in the environment, the isotopologues of the environmental organochlorines are extremely impossible to follow binomial distribution, and the distribution modes may be various depending on the sources of the organochlorines. Therefore, it may be irrational to apply a pair of neighboring isotopologues to calculating the chlorine isotope ratio and regard it as the overall isotope ratio of an organochlorine. It would be better to use all molecular isotopologues for calculating the overall chlorine isotope ratio of an organochlorine.

Source apportionment may not be successfully performed if merely based on the obtained overall chlorine isotope ratios of an organochlorine in two or more samples. Even though the overall chlorine isotope ratios are alike in some cases, the organochlorine in these samples cannot be fully confirmed to stem from the same source, because the organochlorine in different samples could exhibit different distributions of isotopologues. Thus, determination of the isotopologue distribution patterns may be optimal for source apportionment of chlorinated organic compounds.

Using different pairs of adjacent isotopologues may figure out different chlorine isotope ratios (Tang et al., 2017a), of which the distribution modes, the shapes of fitted curves and the fitted equations may be different for the groups of an organochlorine from different sources. To the contrary, if two or more groups of an organochlorine have similar fitted curves and fitted equations, they are very likely from the same source.

As shown in Figure 2, the measured chlorine isotope ratios of perchloroethylene (PCE) and trichloroethylene (TCE) calculated with different pairs of adjacent isotopologues via eq 3.4 are various. The patterns of the isotope ratios measured with the two different types of mass spectrometry were different, which can be illustrated more clearly by the fitted curves and equations as Figure 2 shows. The dechlorination processes in the two mass spectrometers might be different, thus leading to different chlorine KIEs and presenting different changing modes of the isotope ratios calculated with the isotopologue-pair scheme. With the same mass spectrometry (gas chromatography-double focus magnetic-sector high resolution mass spectrometry (GC-DFS-HRMS)), the measured isotope ratios of the two analytes from different manufacturers were different, as well as the fitted curves and equations as shown in Figure 2.

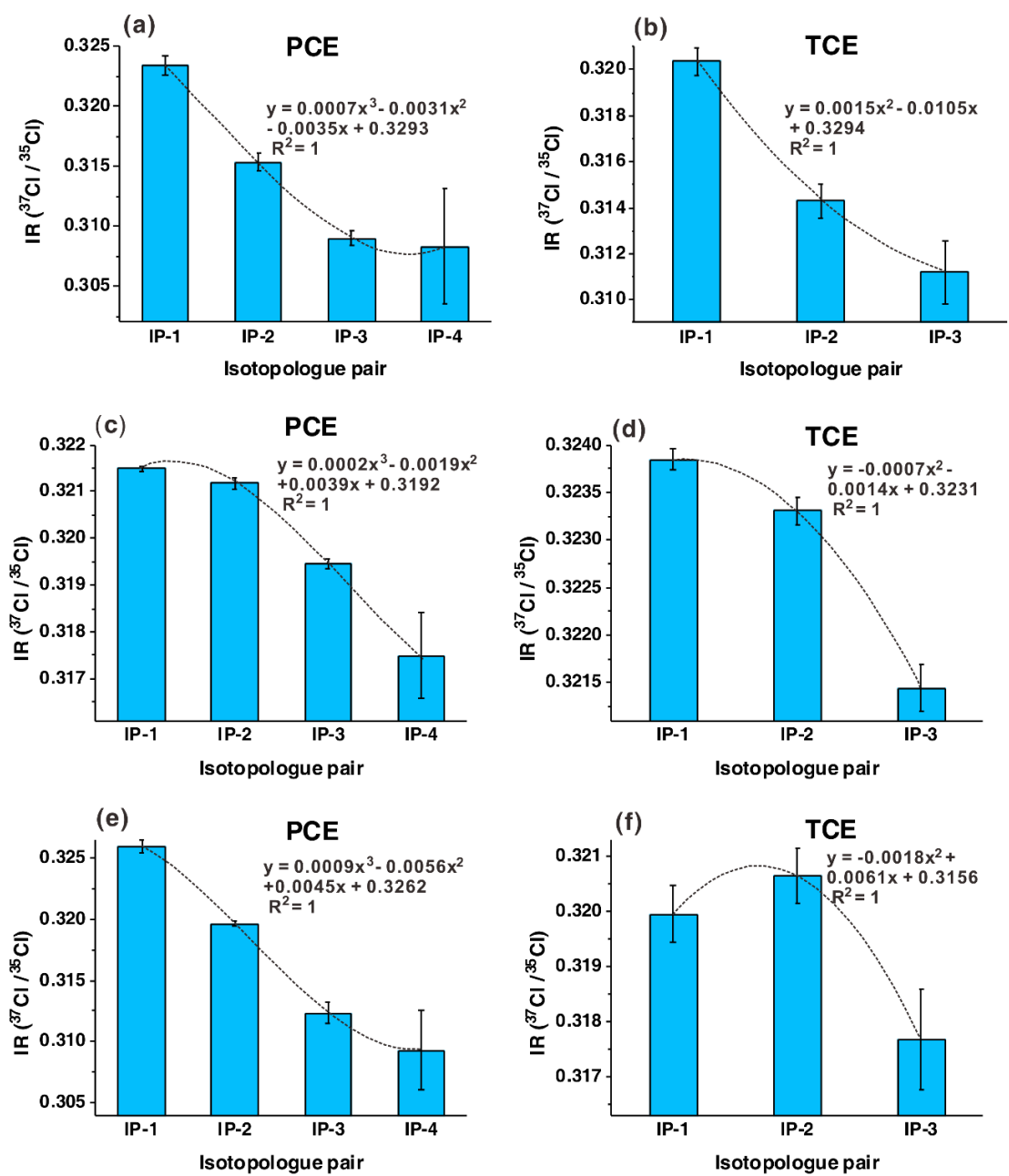

**Figure 2.** Measured chlorine isotope ratios of perchloroethylene (PCE) and trichloroethylene (TCE) calculated with the isotopologue-pair scheme using different pairs of neighboring molecular isotopologues by means of gas chromatography-double focus magnetic-sector high resolution mass spectrometry (GC-DFS-HRMS) or gas chromatography-quadrupole mass spectrometry (GC-qMS). IP: isotopologue pair; IR: isotope ratio. The correlations between the measured chlorine isotope ratios (y) and the numbers of heavy isotope atoms (x) are fitted with polynomial functions. Error bars show the standard deviations (1 σ). The standards PCE and TCE (high performance liquid chromatography grade) in **a**, **b**, **c** and **d** were bought from the manufacturer-1, and those in **e** and **f** were purchased from manufacturer-2 (analytical reagent grade). The injection replicates were five in **a**, and those in others were six.

As Figure 3 and Table S-3 show, the measured chlorine isotope ratios calculated with the first pair of isotopologues were statistically higher than the corresponding isotope ratios calculated with the complete-isotopologue scheme in most cases, except that of the PCE from the manufacturer 2 detected by GC-DFS-HRMS, of which the isotope ratios were statistically indistinguishable. Moreover, the patterns of the isotope ratios calculated with the isotopologue-pair scheme were different from those of the isotope ratios calculated with the complete-isotopologue scheme (Figure 3). Thus, as mentioned previously (Tang et al., 2017a), the standards of the two analytes from any one manufacturer cannot be used as the external isotopic standards for those from another manufacturer when the isotopologue-pair scheme is applied to calculating chlorine isotope ratios.

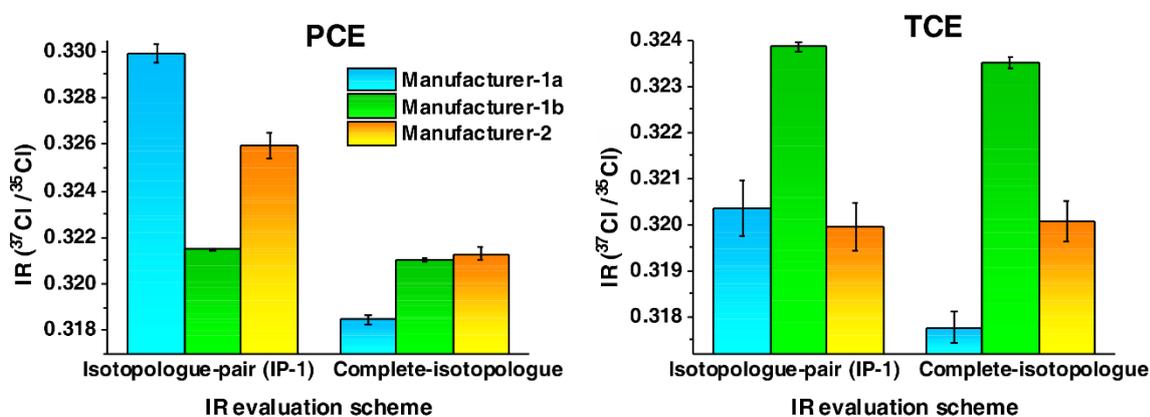

**Figure 3.** Measured chlorine isotope ratios of PCE and TCE from two manufacturers calculated with the first pair of neighboring molecular isotopologues, and those calculated with the complete-isotopologue scheme with the detection of GC-DFS-HRMS or GC-qMS. The standards from different manufacturers were analyzed alternately and successively, and the injection replicates were six. Manufacturer-1b: the corresponding data were obtained with GC-qMS analysis; the rest data were obtained with GC-DFS-HRMS.

Up to now, most reported CSIA-Cl methods involving gas chromatography quadrupole mass spectrometry (GC-qMS) applied the first pair of neighboring isotopologues to calculate the chlorine isotopes ratios (Bernstein et al., 2011; Aeppli et al., 2010; Miska et al., 2015; Heckel et al., 2017). If using the isotopologue-pair scheme to calculate chlorine isotope ratio of a compound and using external isotopic standards which are structurally identical to the target compound for calibration, then the isotopic standards and the target compound should have similar fitted curves of the isotope ratios calculated with different isotopologue pairs versus the sequence numbers of isotopologue pairs. In other words, the fitted curves should be the same except the different Y-intercepts, which means they can overlap after translation parallel with Y-axis (isotope ratio) (Figure 4). In this way, the isotope ratios calculated with the isotopologue-pair scheme can be calibrated to the values equivalent to those calculated with the complete-isotopologue scheme after calibration with the external isotopic standards.

For instance, as shown in Figure 4, the pattern of isotope ratios calculated with the isotopologue-pair scheme of the PCB-18 isotopologues in the first retention-time segment is almost parallel with that in the last one. In addition, the isotope ratio differences of the corresponding isotopologue pairs in the two retention-time segments are close to that of the total isotopologues in the two retention-time segments. Therefore, if we can separately collect the PCB-18 isotopologues of the two retention-time segments, then these two groups of PCB-18 can be used as the external isotopic standards for one another when the isotopologue-pair scheme is applied. The rational external isotopic standards could be prepared with physical separation, such as separation using preparative gas chromatography.

Certainly, we propose the complete-isotopologue scheme to calculate isotope ratios when conducting CSIA-Cl studies. However, this scheme may not be suitable for GC-qMS when analytes have more than two Cl atoms, due to the low signal intensities of the heavier isotopologues. While GC-DFS-HRMS can be applied to CSIA-Cl with the complete-isotopologue scheme for the organochlorines containing no more than six Cl atoms, due to its excellent sensitivity and selectivity.

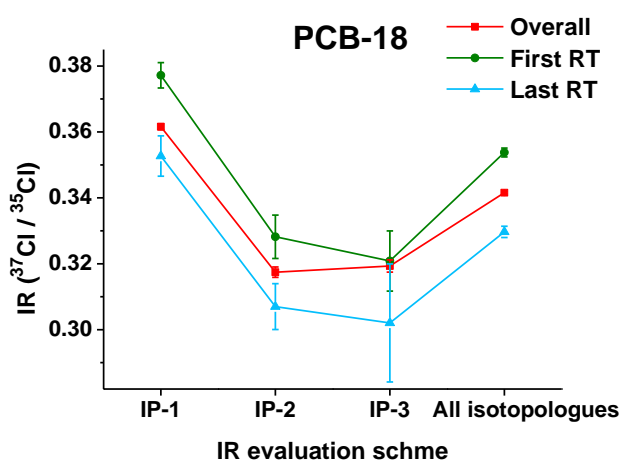

**Figure 4.** Measured chlorine isotope ratios of PCB-18 in different retention-time segments calculated with different pairs of neighboring molecular isotopologues, and those calculated with complete isotopologues with the detection of GC-DFS-HRMS. Overall: the isotope ratios

in the whole chromatographic peak; First RT: the isotope ratios in the first retention-time segment; Last RT: the isotope ratios in the last retention-time segment.

## CONCLUSIONS

This study provides a solid theoretical inference for proving whether the chlorine isotopologues of organochlorines are binomially distributed or not, and explores the impacts of the distributions on the quantification and CSIA-Cl of organochlorines. The fundamentals that cause different chlorine KIEs were discussed. During synthetic reactions, the C-Cl bonds having stronger strengths are more likely to be formed with heavy isotopes. The stronger C-Cl bonds exhibit higher chlorine KIEs during breaking comparing with the weaker bonds. For synthetic organochlorines, if the rate-limiting steps in the chlorination reactions are the formations of C-Cl bonds, then chlorine KIEs work during the reactions, causing always higher chlorine isotope ratios on the reaction positions than the initial isotope ratios of the chlorine sources. Chlorine KIEs are different during different chlorination processes, resulting in inequivalent chlorine isotope ratios on different reaction positions of asymmetric reaction intermediates and final products. If the rate-determining step is the production process of reactive chlorine atoms and the initial isotope ratio of the chlorine source keeps changeless, then the chlorine isotope ratios on the reaction positions of the intermediates and final products are equal, indicating the isotopologues complying with binomial distribution. After physical changes and dechlorination reactions in the environment, organochlorines are impossibly binomially distributed no matter what the initial isotopologue distributions are. Experimental results show that the detected isotopologues of all the investigated organochlorines do not follow binomial distribution. Different isotopologue distributions may trigger deviations in quantification and CSIA-Cl of organochlorines. Utilization of more than one highest-abundance isotopologue for quantification of organochlorines and application of complete-isotopologue scheme of isotope ratio evaluation for CSIA-Cl are proposed for achieving high-quality results. GC-DFS-HRMS may be promising to perform CSIA-Cl using the complete-isotopologue scheme, owing to the excellent sensitivity and selectivity. This study may update the recognition of people with regard to the chlorine isotopologue distributions of organochlorines, and the proposed solutions may be helpful to get high-quality experimental results in quantification and CSIA-Cl studies for organochlorines.

## ASSOCIATED CONTENT

The *Supporting Information* containing the experimental details and results is available free of charge on the website at http://pending.

## ACKNOWLEDGEMENTS

The author is grateful for Mr. Jianhua Tan and Mr. Deyun Liu, from Guangzhou Quality Supervision and Testing Institute, China for their gift of the chloroethylene standards and help in GC-qMS analysis. This work was partially supported by the National Natural Science Foundation of China (Grant No. 41603092).

**Figure Legends**

**Figure 1**. Energy level diagram of reactions involving chlorine isotopes. The zero point energy of a molecule formed by the heavy isotope (ZPE*) is lower than that of the molecule formed with the light one. The extent of ΔZPE (ZPE-ZPE*) for a compound play a major role on the kinetic isotope effect (KIE). The chemical bonds containing heavy isotopes are faster to be formed but slower to be cleaved than those containing light isotopes. $\Delta G^{\neq}$ represents the critical energies; R is the universal gas constant (0.00199 kcal/mol·K); T is temperature (K).

**Figure 2.** Measured chlorine isotope ratios of perchloroethylene (PCE) and trichloroethylene (TCE) calculated with the isotopologue-pair scheme using different pairs of neighboring molecular isotopologues by means of gas chromatography-double focus magnetic-sector high resolution mass spectrometry (GC-DFS-HRMS) or gas chromatography-quadrupole mass spectrometry (GC-qMS). IP: isotopologue pair; IR: isotope ratio. The correlations between the measured chlorine isotope ratios (y) and the numbers of heavy isotope atoms (x) are fitted with polynomial functions. Error bars show the standard deviations (1 σ). The standards PCE and TCE (high performance liquid chromatography grade) in **a**, **b**, **c** and **d** were bought from the manufacturer-1, and those in **e** and **f** were purchased from manufacturer-2 (analytical reagent grade). The injection replicates were five in **a**, and those in others were six.

**Figure 3.** Measured chlorine isotope ratios of PCE and TCE from two manufacturers calculated with the first pair of neighboring molecular isotopologues, and those calculated with the complete-isotopologue scheme with the detection of GC-DFS-HRMS or GC-qMS. The standards from different manufacturers were analyzed alternately and successively, and the injection replicates were six. Manufacturer-1b: the corresponding data were obtained with GC-qMS analysis; the rest data were obtained with GC-DFS-HRMS.

**Figure 4.** Measured chlorine isotope ratios of PCB-18 in different retention-time segments calculated with different pairs of neighboring molecular isotopologues, and those calculated with complete isotopologues with the detection of GC-DFS-HRMS. Overall: the isotope ratios in the whole chromatographic peak; First RT: the isotope ratios in the first retention-time segment; Last RT: the isotope ratios in the last retention-time segment.

**Table Caption**

**Table 1.** Quantification biases calculated with the measured relative abundances of different isotopologues of PCB-18 in the first and the last retention-time segments.